\documentclass[twocolumn,times]{aastex63}
\usepackage[T1]{fontenc}
\usepackage{enumitem}
\usepackage{paralist}

\newcommand\ergs{erg~s$^{-1}$}
\newcommand\ergcms{erg~cm$^{-2}$~s$^{-1}$}

\newcommand{\HII}{H\,{\sc ii}}

\defcitealias{She2017_PaperI}{Paper I}
\defcitealias{She2017_PaperII}{Paper II}
\defcitealias{She2018_PaperIII}{Paper III}

\shorttitle{Chandra Survey of Nearby Galaxies}
\shortauthors{Bi et al.}

\begin{document}
\title{Chandra Survey of Nearby Galaxies: an Extended Catalog}

\author{Sheng Bi}
\affiliation{Department of Physics, Tsinghua University, Beijing 100084, China}

\author[0000-0001-7584-6236]{Hua Feng}
\affiliation{Department of Astronomy, Tsinghua University, Beijing 100084, China}
\affiliation{Department of Engineering Physics, Tsinghua University, Beijing 100084, China}

\author{Luis C.\ Ho}
\affiliation{Kavli Institute for Astronomy and Astrophysics, Peking University, Beijing 100087, China}
\affiliation{Department of Astronomy, Peking University, Beijing 100087, China}

\correspondingauthor{Hua Feng}
\email{hfeng@tsinghua.edu.cn}

\tabletypesize{\scriptsize}

\begin{abstract}

She et al.~(Paper I) assembled a catalog of nearby galaxies observed with the Chandra X-ray observatory, by cross-matching galaxies in the NASA Extragalactic Database (NED) within 50~Mpc and the Chandra archive. That sample has enabled searches of low-mass black holes associated with late-type, bulgeless galaxies and studies of the accretion physics related to low-luminosity active galactic nuclei (LLAGNs).  Using a similar approach, here we construct an extended catalog up to 150~Mpc and make a cross-correlation with a catalog of nearby galaxy groups.  The new catalog consists of 1,964 galaxies, out of which 1,692 have a redshift independent distance, 1,557 are listed in the galaxy group catalog with group properties available, and 782 are identified to be X-ray AGN candidates. Compared with the AGN sample in Paper I, the new sample is 2.5 times larger in size (782 vs.\ 314), with $\sim$80\% of the new members having an Eddington ratio less than $10^{-4}$.  We confirm that the conclusions based on the previous sample remain. With the new sample, we compare AGN fractions between early-type and late-type galaxies, and between central and satellite galaxies in groups, and find no significant difference. This suggests that the secular process is not the dominant mechanism feeding AGNs in the local universe. 

\end{abstract}


\section{Introduction}
\label{sec:intro}

Powered by accretion onto supermassive black holes (SMBHs), active galactic nuclei (AGNs) are an important ingredient in the growth and evolution of their host galaxies \citep[for a review, see][]{Kormendy2013}.  A sample of AGNs in nearby galaxies may help address or shed light on three intriguing questions: how were the SMBHs formed in the early universe \citep{Volonteri2010}, what is the physics of accretion when the accretion rate is extremely low \citep{Yuan2014}, and what is the feeding mechanism for SMBHs in the nearby universe \citep{Storchi-bergmann2019}. For the first two questions, \citet[][\citetalias{She2017_PaperI}]{She2017_PaperI} assembled a catalog of Chandra observed galaxies within 50~Mpc and identified 314 candidate AGNs via X-ray observations. They argued that X-rays are more powerful and robust than the optical band in revealing weak AGN activities, thanks to less contamination from stellar processes in the host galaxy.

Based on that AGN sample, \citet[][\citetalias{She2017_PaperII}]{She2017_PaperII} found that $\sim$30\% of the \HII\ nuclei contain an AGN. These \HII\ nuclei predominantly belong to late-type, bulgeless galaxies, which are expected to harbor low-mass black holes, if any, according to the galaxy-black hole scaling law \citep{Kormendy2013}. The occupation fraction of low-mass black holes in these galaxies can place a stringent constraint on the formation mechanism of the SMBHs in the early universe \citep{Greene2020}. Taking into account possible contaminations from X-ray binaries, they estimated a fraction of at least 21\% of low-mass black holes in late-type galaxies.

A significant fraction of the AGNs in the sample are LLAGNs, in which both the accretion rate and efficiency are low and a hot accretion flow is expected \citep[for a review, see][]{Yuan2014}. \citet[][\citetalias{She2018_PaperIII}]{She2018_PaperIII} revealed a positive correlation between the Eddington ratio and the X-ray absorption column density (or similarly, the number of highly absorbed AGNs). This suggests that X-ray absorption originates in or is correlated with the outflow of the hot accretion flow, in good agreement with numerical simulations \citep{Yuan2015}.

Another interesting question is about the triggering and feeding mechanisms of SMBHs \citep[for a review, see][]{Storchi-bergmann2019}, which could be heuristic to study the evolution history of SMBHs and their regulation of the environment. Possible mechanisms include galaxy mergers, chaotic cold accretion \citep{Gaspari2013} on extragalactic scales, and secular process \citep{Kormendy2004} on galactic scales.  To investigate this problem, a relatively large sample is needed, as well as environmental information given a galaxy.  

To enlarge the sample size, here we follow the procedures in \citetalias{She2017_PaperI} and extend the search to a distance of 150~Mpc. In addition, we adopt the galaxy group catalog of \citet{Tully2015} to get a distance estimate for galaxies without a direct distance\footnote{distance not derived from redshift.} measurement.  In those cases, the distance of other group members or the distance converted from the peculiar velocity-corrected redshift is adopted.  The group catalog also provides environmental information of galaxies. As a result, the new catalog includes 1,964 galaxies, among which 782 AGN candidates are identified.  The sample construction and data reduction is described in \S~\ref{sec:sample}. The results are presented in \S~\ref{sec:result} and discussed in \S~\ref{sec:discussion}. 

\section{Sample}
\label{sec:sample}

The search is based on Chandra ACIS observations available in the public archive as of October in 2018, by cross-matching galaxies in the NASA/IPAC Extragalactic Database\footnote{http://ned.ipac.caltech.edu} (NED) within 150~Mpc, with the galaxy group catalog of \citet{Tully2015} as a supplement.  The processes used here are generally adopted from those in \citetalias{She2017_PaperI}, with a few modifications.  The readers may refer to \citetalias{She2017_PaperI} for more technical details if they want to repeat our work. Otherwise, the detail level in this paper should be sufficient for readers to understand the purpose and logic of each step.

Galaxies with an angular separation less than 8\arcmin\ to the aim point of any observation are checked with the \textit{dmcoords} task in CIAO to see if they are located in the Chandra field of view. The distance to the galaxy is adopted from NED, with a defined order of priority; the most recent reference is adopted if there is more than one entry available in the same category of distance measurements/estimates.  This part is identical to that in \citetalias{She2017_PaperI}.  The \citet{Tully2015} galaxy group catalog, constructed from the 2MASS Redshift Survey with almost 91\% completeness over the sky to $K_s = 11.75$, provides supplementary information to the distance.  If a galaxy has no direct distance measurement/estimate, we adopt the distance of the central galaxy of the group, or the mean distance of the rest of the members if the central galaxy has no direct distance. Otherwise, if none of the group members has a direct distance, the group recession velocity, which has been corrected for peculiar motions, is translated to distance assuming a Hubble constant $h = 0.7$.

We extend the search to a distance of 150~Mpc, or a recession velocity of about 10,000~km~s$^{-1}$ \citep[above which the group properties are considered unreliable;][]{Tully2015}. As a result, we collect 1,478 galaxies with a direct distance available for themselves, 214 galaxies with a direct distance from their group members, and 272 galaxies with an indirect distance derived from the group velocity.  Among them, 1,557 galaxies are quoted in \citet{Tully2015}, constituting an interesting sub-sample with useful group properties.  Due to the update of NED, some galaxies may have a new distance here compared with those in \citetalias{She2017_PaperI}.

The stellar mass of the host galaxy is estimated from the $K_s$ band luminosity \citep{Skrutskie2006} assuming a constant mass-to-light ratio, which is derived as follows.  Following \citet{Bell2001}, the stellar mass can be expressed as a function of the $K_s$ band luminosity corrected by the \bv\ color. We adopt the coefficients assuming a scaled \citet{Salpeter1955} initial mass function (IMF) and $Z = 0.02$, and subtract a zero point of 0.15 dex \citep{Bell2003} to transfer from the scaled Salpeter IMF to \citet{Kroupa2001} IMF, which is close to the \citet{Chabrier2001} IMF \citep{Madau2014}. This gives a mass estimation of
\begin{equation}
\log(M / M_{\odot}) = \log(L_{K}/L_{\odot}) + 0.58(B-V) - 0.20 \; .
\label{eq:mass}
\end{equation}
The $K_s$ band luminosity can be found for 1,557 objects in our sample, while more than half of them do not have a \bv\ color. For objects in our sample, the above estimate gives a median $M/L = 0.63$.  A stellar mass derived using such a simple mass-to-light ratio deviates from the mass inferred using Equation~(\ref{eq:mass}) by a median of 0.1~dex, smaller than the intrinsic scatter itself \citep{Bell2001}. Thus, a constant mass-to-light ratio of 0.63 is adopted in this work to estimate the stellar mass from the $K_s$ band luminosity.

The star formation rate (SFR) of the host galaxy is estimated from the infrared flux measured with the Infrared Astronomical Satellite (IRAS), following the recipe in \citet{Kennicutt1998} as
\begin{equation}
{\rm SFR} / (M_\odot\; {\rm yr}^{-1}) = 3.0 \, L_{\rm FIR} / (10^{44}\; {\rm erg\; s}^{-1}),
\end{equation}
where $L_{\rm FIR}$ is computed from the 60~$\mu$m and 100~$\mu$m flux density following \citet{Helou1985}, and the Salpeter IMF has been transferred to the Kroupa IMF \citep{Madau2014}.

The X-ray data reduction processes are identical to those in \citetalias{She2017_PaperI}. The longest exposure is used if there are multiple observations of the same galaxy. CIAO 4.9 with CALDB 4.7.3 is used to reduce the Chandra data. New events files are created using the task \textit{chandra\_repro}. Time intervals with high background are excluded using \textit{lc\_clean} for moderate flares or \textit{lc\_sigma\_clip} with $3\sigma$ clipping for heavy flares. Then, X-ray images and exposure maps are created using \textit{fluximage} assuming a power-law weighting spectrum with a photon index $\Gamma=1.8$ \citepalias[typical for nearby AGNs;][]{She2017_PaperI} subject to Galactic absorption. Sources are detected using \textit{wavdetect} with point spread function (PSF) maps generated using \textit{mkpsfmap} at the mean energy of the weighting spectrum. Source regions that encircle 98.9\% (3$\sigma$ of a 2D Gaussian) of the PSF are used to extract source photons. In order to remove fake or extended sources, we only select objects that have at least 7 photons, a significance higher than 3, and a PSF ratio less than 3; these criteria were not imposed in \citetalias{She2017_PaperI}. As a result, 18 objects in \citetalias{She2017_PaperI} no longer appear in the new catalog.

Given the near-infrared/optical center of a galaxy, the identification of an AGN candidate is based on the test that an X-ray source is spatially consistent with the galaxy center within errors, including their statistical errors and the absolute astrometry of Chandra, at 99\% confidence level.  For each AGN candidate, a background region (5\arcsec\ to 10\arcsec\ annulus, or a nearby circular region if there is source confusion) is defined to calculate the background-corrected count rate and hardness ratio, using the task \textit{aprates} that accounts for Poisson fluctuations in manner of Bayesian approach.  X-ray spectra are extracted using \textit{specextract} if there are at least 100 photons in 0.3--8 keV. Adopting the same approaches used in \citetalias{She2017_PaperI}, a single {\tt powerlaw} model or a two-component model ({\tt powerlaw} plus {\tt mekal} or {\tt bbody}) is used to fit the spectra in XSPEC.  The energy spectrum can also help discriminate fake identifications from confusion with emission from central hot gas.  If {\tt mekal} is the only model that provides an acceptable fit, we remove the object from our catalog. If {\tt mekal} and a non-thermal model both provide acceptable fits,  we check the radial profile and remove objects that are are wider than the local PSF. The observed flux and intrinsic luminosity in 2--10 keV are then calculated from the best-fit spectra using \textit{cflux} or translated from count rates assuming an absorbed power-law model with local response files.  The Eddington ratio ($\lambda_{\rm Edd}$) is obtained from the X-ray luminosity assuming a bolometric correction factor of 16, obtained from broadband SEDs of LLAGNs \citep{Ho2008}.  

The black hole mass is estimated via the $M_{\rm BH}-\sigma_\ast$ relation in \citetalias{She2017_PaperI}. However, central stellar velocity dispersions are available only for 58\% of the AGN candidates in our sample.  Here, we adopt the scaling between the black hole mass and the total stellar mass ($M_\ast$) of the galaxy to infer $M_{\rm BH}$ using the recipe quoted in \citet{Greene2020}, in particular, the one separated for early and late-type galaxies with upper limits used for the latter.  This calibration is found to have the smallest intrinsic scatter of about 0.65~dex \citep{Greene2020}, while the intrinsic scatter of the $M_{\rm BH}-\sigma_\ast$ relation used in \citetalias{She2017_PaperI} is 0.44~dex. The $M_{\rm BH}$ inferred from $\sigma_\ast$ vs.\ that from $M_\ast$ for objects in our sample has a scatter of 0.8~dex.  Thus, this approach introduces an acceptable uncertainty in $M_{\rm BH}$, but allows us to evaluate the results for the whole sample, and is adopted in the following work. In addition to \citetalias{She2017_PaperI}, we calculate the limiting luminosity ($L_{\rm limit}$) at a confidence level of 3$\sigma$ for each observation at the position of galactic nucleus, based on the background rate and PSF size, assuming the same power-law model as mentioned above.  

The properties of galaxies and AGNs are tabulated in a machine readable table available on line, see an example in Table~\ref{tab:sample} with explanations of table columns. The X-ray properties for AGN candidates already in \citetalias{She2017_PaperI} are not reprocessed except for the luminosity if there is an update of the distance.  

\begin{deluxetable*}{llllllllll}
\tablewidth{\textwidth}
\tablecaption{Properties of galaxies and AGNs in our sample \label{tab:sample}}
\tablehead{
\colhead{ID} & \colhead{name} & \colhead{PGC} & \colhead{group flag} & \colhead{$d$} & \colhead{$d_{\rm direct}$} & \colhead{$d_{\rm mean}$} & \colhead{$d_{\rm group}$} & \colhead{R.A.} & \colhead{Decl.} \\ 
\colhead{} & \colhead{} & \colhead{} & \colhead{} & \colhead{(Mpc)} & \colhead{(Mpc)} & \colhead{(Mpc)} & \colhead{(Mpc)} & \colhead{(J2000)} & \colhead{(J2000)} \\
\colhead{(1)} & \colhead{(2)} & \colhead{(3)} & \colhead{(4)} & \colhead{(5)} & \colhead{(6)} & \colhead{(7)} & \colhead{(8)} & \colhead{(9)} & \colhead{(10)} \\
\noalign{\smallskip}\hline\noalign{\smallskip}
\colhead{class} & \colhead{Hubble type} & \colhead{NUV$- B$} & \colhead{$\log \frac{M_{\rm gal}}{M_\sun}$} & \colhead{$\log\frac{\rm SFR}{M_\sun \; {\rm yr}^{-1}}$} & \colhead{$\sigma_{\ast}$} & \colhead{$\sigma_{\ast, \rm err}$} & \colhead{$\log\frac{L_{\rm H\alpha}}{{\rm erg\; s}^{-1}}$} & \colhead{note on $L_{\rm H\alpha}$} & \colhead{$\log\frac{M_{\rm BH}}{M_\sun}$} \\
\colhead{} & \colhead{} & \colhead{(mag)} & \colhead{} & \colhead{} & \colhead{(km s$^{-1}$)} & \colhead{(km s$^{-1}$)} & \colhead{} & \colhead{} & \colhead{} \\
\colhead{(11)} & \colhead{(12)} & \colhead{(13)} & \colhead{(14)} & \colhead{(15)} & \colhead{(16)} & \colhead{(17)} & \colhead{(18)} & \colhead{(19)} & \colhead{(20)} \\
\noalign{\smallskip}\hline\noalign{\smallskip}
\colhead{$\log\frac{M_{\rm BH,lo}}{M_\sun}$} & \colhead{$\log\frac{M_{\rm BH,up}}{M_\sun}$} & \colhead{$\log\frac{M_{\rm BH}^\ast}{M_\sun}$} & \colhead{$\log\frac{M_{\rm BH,lo}^\ast}{M_\sun}$} & \colhead{$\log\frac{ M_{\rm BH,up}^\ast}{M_\sun}$} & \colhead{$N_{\rm H, Gal}$} & \colhead{ID in Paper I} & \colhead{ObsID} & \colhead{instrument} & \colhead{exposure} \\
\colhead{} & \colhead{} & \colhead{} & \colhead{} & \colhead{} & \colhead{($10^{22}$~cm$^{-2}$)} & \colhead{} & \colhead{} & \colhead{} & \colhead{(ks)} \\
\colhead{(21)} & \colhead{(22)} & \colhead{(23)} & \colhead{(24)} & \colhead{(25)} & \colhead{(26)} & \colhead{(27)} & \colhead{(28)} & \colhead{(29)} & \colhead{(30)} \\
\noalign{\smallskip}\hline\noalign{\smallskip}
 \colhead{offset} & \colhead{AGN} & \colhead{$H_0$} & \colhead{$H_{0,\rm err}$} & \colhead{$H_1$} & \colhead{$H_{1,\rm err}$} & \colhead{$H_2$} & \colhead{$H_{2,\rm err}$} & \colhead{$N_{\rm H,HR}$} & \colhead{$N_{\rm H,HR,lo}$} \\
 \colhead{($^{\prime\prime}$)} & \colhead{} & \colhead{} & \colhead{} & \colhead{} & \colhead{} & \colhead{} & \colhead{} & \colhead{($10^{22}$~cm$^{-2}$)} & \colhead{($10^{22}$~cm$^{-2}$)} \\
\colhead{(31)} & \colhead{(32)} & \colhead{(33)} & \colhead{(34)} & \colhead{(35)} & \colhead{(36)} & \colhead{(37)} & \colhead{(38)} & \colhead{(39)} & \colhead{(40)} \\
\noalign{\smallskip}\hline\noalign{\smallskip}
\colhead{$N_{\rm H,HR,up}$} & \colhead{$f_{\rm X}$} & \colhead{$f_{\rm X,lo}$} & \colhead{$f_{\rm X,up}$} & \colhead{$\log\frac{L_{\rm lim} }{{\rm erg\; s}^{-1}}$} & \colhead{$\log\frac{L_{\rm X} }{{\rm erg\; s}^{-1}}$} & \colhead{$\log\frac{L_{\rm X,lo} }{{\rm erg\; s}^{-1}}$} & \colhead{$\log\frac{L_{\rm X,up}}{{\rm erg\; s}^{-1}}$} & \colhead{$\lambda_{\rm Edd}$} & \colhead{$\lambda_{\rm Edd,lo}$} \\
\colhead{($10^{22}$~cm$^{-2}$)} & \colhead{($10^{-14}$~cgs)} & \colhead{($10^{-14}$~cgs)} & \colhead{($10^{-14}$~cgs)} & \colhead{} & \colhead{} & \colhead{} & \colhead{} & \colhead{} & \colhead{} \\
\colhead{(41)} & \colhead{(42)} & \colhead{(43)} & \colhead{(44)} & \colhead{(45)} & \colhead{(46)} & \colhead{(47)} & \colhead{(48)} & \colhead{(49)} & \colhead{(50)} \\
\noalign{\smallskip}\hline\noalign{\smallskip}
\colhead{$\lambda_{\rm Edd,up}$} & \colhead{model} & \colhead{$\log$ Norm$_1$} & \colhead{$\log\frac{ L_1 }{{\rm erg\; s}^{-1}}$} & \colhead{$\log$ Norm$_2$} & \colhead{$\log\frac{L_2}{{\rm erg\; s}^{-1}}$} & \colhead{$N_{\rm H}$} & \colhead{$N_{\rm H,lo}$} & \colhead{$N_{\rm H,up}$} & \colhead{$\Gamma$} \\
\colhead{} & \colhead{} & \colhead{} & \colhead{} & \colhead{} & \colhead{} & \colhead{($10^{22}$ cm$^{-2}$)} & \colhead{($10^{22}$ cm$^{-2}$)} & \colhead{($10^{22}$ cm$^{-2}$)} & \colhead{}\\
\colhead{(51)} & \colhead{(52)} & \colhead{(53)} & \colhead{(54)} & \colhead{(55)} & \colhead{(56)} & \colhead{(57)} & \colhead{(58)} & \colhead{(59)} & \colhead{(60)} \\
\noalign{\smallskip}\hline\noalign{\smallskip}
\colhead{$\Gamma_{\rm lo}$} &  \colhead{$\Gamma_{\rm up}$} & \colhead{$T$} & \colhead{$T_{\rm lo}$} & \colhead{$T_{\rm up}$} & \colhead{$\chi^2$} & \colhead{d.o.f.} & \colhead{note on $L_{\rm X}$} & \colhead{} & \colhead{} \\
\colhead{} & \colhead{} & \colhead{(keV)} & \colhead{(keV)} & \colhead{(keV)} & \colhead{} & \colhead{} & \colhead{} & \colhead{} & \colhead{} \\
\colhead{(61)} & \colhead{(62)} & \colhead{(63)} & \colhead{(64)} & \colhead{(65)} & \colhead{(66)} & \colhead{(67)} & \colhead{(68)} & \colhead{} & \colhead{} 
}
\startdata
\enddata
\tablecomments{Column 1: object ID.
Column 2: common name of the galaxy.
Column 3: PGC number of the galaxy.
Column 4: in \citet{Tully2015} or not.
Column 5: distance used in the catalog.
Column 6: direct distance.
Column 7: mean, direct distance of the group members.
Column 8: distance converted from the group recession velocity assuming $h = 0.7$.
Column 9: right ascensionof the galaxy center.
Column 10: declination of the galaxy center.
Column 11: nuclear spectral classification, which is adopted from various sources \citep{Ho1997,Veron-cetty2010,Moustakas2006} or calculated from optical spectra \citep{Kennicutt1992,Falco1999,Colless2003,Jones2009,Rosales-ortega2010,Driver2011}, following the protocol elaborated in \citetalias{She2017_PaperI}.
Column 12: Hubble type quoted in NED.
Column 13:  Galactic extinction \citep{Cardelli1989} corrected color, where NUV is the Kron-like elliptical aperture magnitude of the host galaxy adopted from GALEX \citep{Bianchi2017}, and the $B$ magnitude is adopted from RC3 \citep{deVaucouleurs1991}.
Column 14: stellar mass of the host galaxy.
Column 15: SFR of the host galaxy.
Column 16: central stellar velocity dispersion. In our sample, 920 galaxies have a published $\sigma_\ast$: 292 from the Palomar survey \citep{Ho2009a}, 618 from the HyperLeda database (http://leda.univ-lyon1.fr), and 10 from \citet{Gu2006}. $\sigma_\ast$ in km~s$^{-1}$.
Column 17: error of $\sigma_\ast$ in km~s$^{-1}$.
Column 18: nuclear H$\alpha$ luminosity with data from \citet{Ho1997} and \citet{Ho2003}.
Column 19: note on the H$\alpha$ luminosity.
Column 20: black hole mass calculated assuming the $M_{\rm BH} - \sigma_\ast$ relation with the recipe in \citetalias{She2017_PaperI}.
Column 21: lower limit of the black hole mass.
Column 22: upper limit of the black hole mass.
Column 23: black hole mass inferred using the $M_{\rm BH} - M_{\ast}$ relation with the recipe in \citet{Greene2020}.
Column 24: lower limit of the black hole mass from the $M_{\rm BH} - M_{\ast}$ relation.
Column 25: upper limit of the black hole mass from the $M_{\rm BH} - M_{\ast}$ relation.
Column 26: Galactic absorption column density along the line of sight.
Column 27: object ID in \citetalias{She2017_PaperI} if available.
Column 28: Chandra observation ID.
Column 29: Chandra instrument used for the observation.
Column 30: Chandra exposure time.
Column 31: offset between the X-ray position and the galaxy center.
Column 32: an AGN candidate or not.
Column 33: hardness ratio $H_0 \equiv (C_{\rm H}-C_{\rm M}-C_{\rm S})/C_{\rm T}$, where $C_{\rm S}$, $C_{\rm M}$,  $C_{\rm H}$, and $C_{\rm T}$ are counts in the soft band (0.3$-$1~keV), medium band (1$-$2~keV), hard band (2$-$8~keV), and full band (0.3$-$8~keV), respectively.
Column 34: error of $H_0$.
Column 35: hardness ratio $H_1 \equiv (C_{\rm M}-C_{\rm S})/C_{\rm T}$.
Column 36: error of $H_1$.
Column 37: hardness ratio $H_2 \equiv (C_{\rm H}-C_{\rm M})/C_{\rm T}$.
Column 38: error of $H_2$.
Column 39: interstellar absorption column density beyond the Milky Way, estimated from the hardness ratios assuming a power-law spectrum with a photon index of 1.8.
Column 40: lower limit of $N_{\rm H,HR}$.
Column 41: upper limit of $N_{\rm H,HR}$.
Column 42: observed flux in 2-10 keV in units of $10^{-14}$~\ergcms.
Column 43: lower limit of the observed flux.
Column 44: upper limit of the observed flux.
Column 45: limiting luminosity given the target and observation in 2-10 keV.
Column 46: X-ray luminosity in 2-10 keV after correction for absorption.
Column 47: lower limit of $\log L_{\rm X}$.
Column 48: upper limit of $\log L_{\rm X}$.
Column 49: Eddington ratio $\log (\lambda_{\rm Edd} \equiv 16 L_{\rm X} / L_{\rm Edd})$, where $L_{\rm Edd} = 1.26 \times 10^{38}  (M_{\rm BH} / M_\sun) $~\ergs\ is the Eddington luminosity and 16 is the bolometric correction factor.
Column 50: lower limit of $\log \lambda_{\rm Edd}$.
Column 51: upper limit of $\log \lambda_{\rm Edd}$.
Column 52: best-fit model in XSPEC.
Column 53: normalization of the power-law component in units of photons~cm$^{-2}$~s$^{-1}$~keV$^{-1}$ at 1 keV.
Column 54: luminosity of the power-law component in 0.3-10 keV.
Column 55: normalization (in unit defined in XSPEC) of the second component ({\tt blackbody}/{\tt mekal}).
Column 56: luminosity of the second component in 0.3-10 keV.
Column 57: interstellar absorption column density beyond the Milky Way, derived from spectral fitting.
Column 58: lower limit of $N_{\rm H}$.
Column 59: upper limit of $N_{\rm H}$.
Column 60: power-law photon index.
Column 61: lower limit of $\Gamma$.
Column 62: upper limit of $\Gamma$.
Column 63: temperature of the {\tt mekal} or {\tt blackbody} component.
Column 64: lower limit of the temperature.
Column 65: upper limit of the temperature.
Column 66: $\chi^2$ from the best-fit.
Column 67: degree of freedom of the fit.
Column 68: how the luminosity is obtained: derived from spectral fitting or converted from count rate? If from count rate, is the spectral shape estimated from the hardness ratios or assumed to be a power-law spectrum ($\Gamma = 1.8$:subject to Galactic absorption? Or the reference from which the luminosity is adopted if there is uncorrectable pileup.}

\end{deluxetable*}

\begin{figure}
\centering
\includegraphics[width=\columnwidth]{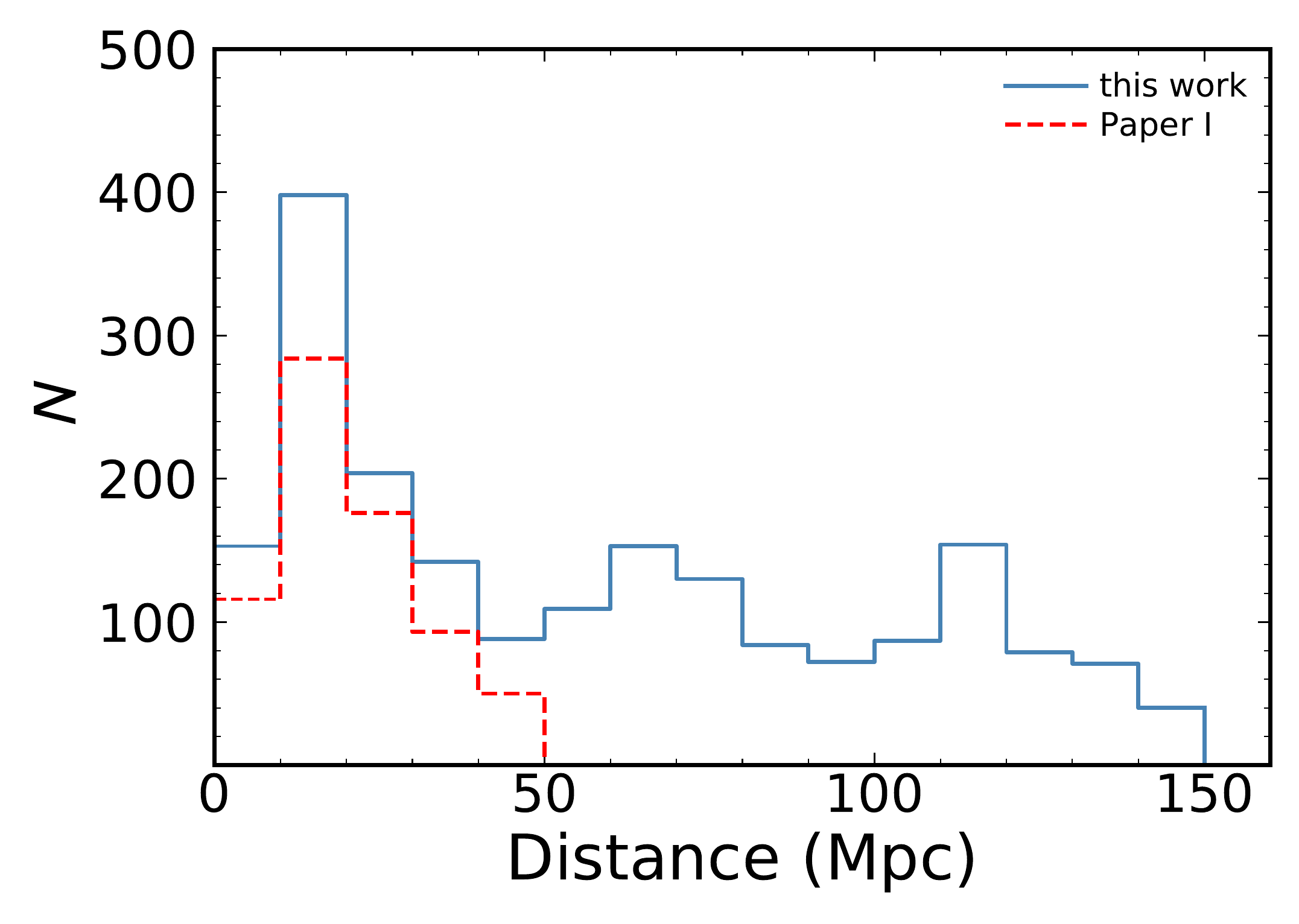}
\caption{Distributions of distance for objects in our sample and in \citetalias{She2017_PaperI}. }
\label{fig:dist}
\end{figure}

\begin{figure}
\centering
\includegraphics[width=\columnwidth]{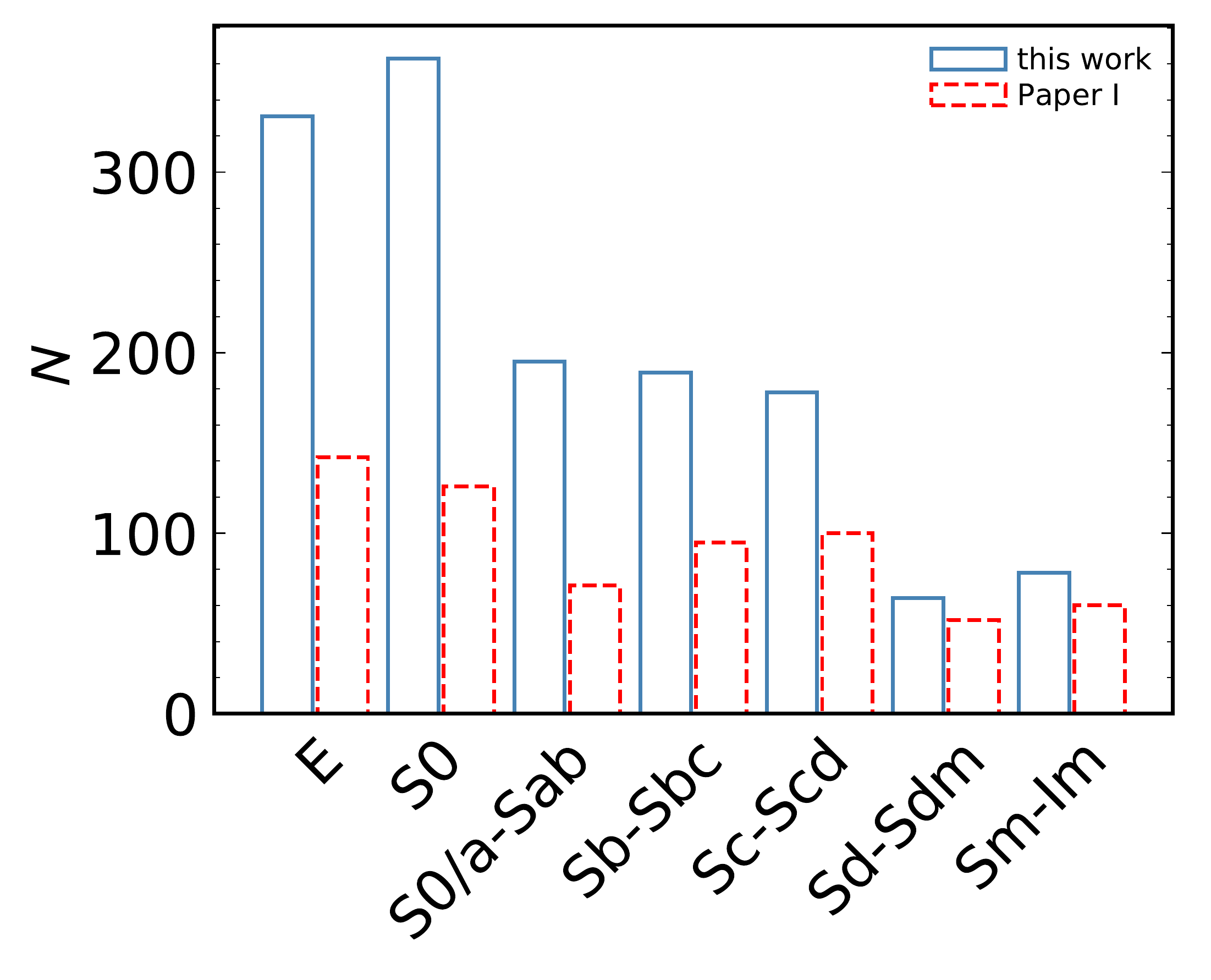}
\caption{Distributions of Hubble types for objects in our sample and in \citetalias{She2017_PaperI}.}
\label{fig:type}
\end{figure}

\begin{figure}
\centering
\includegraphics[width=\columnwidth]{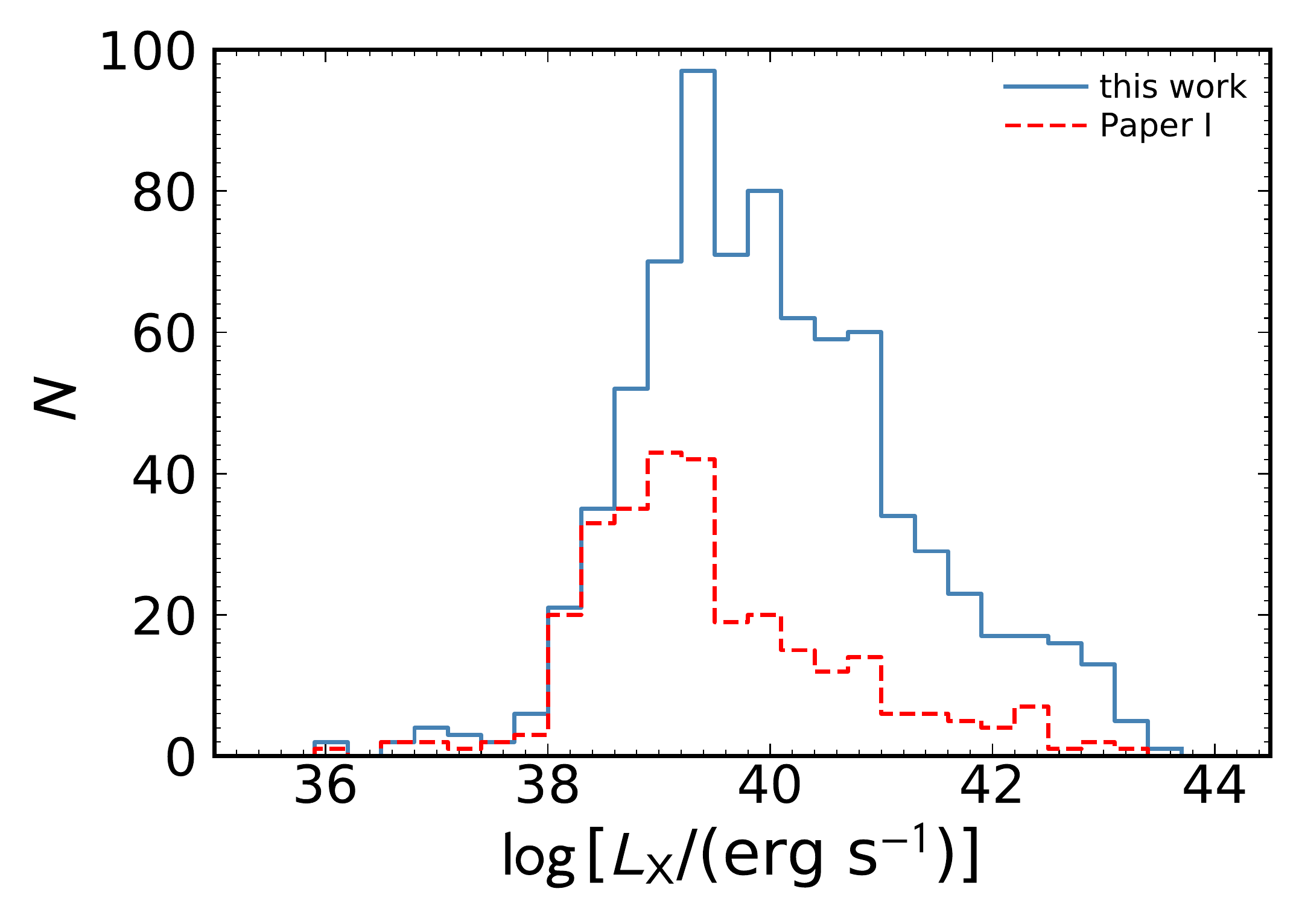}\\
\includegraphics[width=\columnwidth]{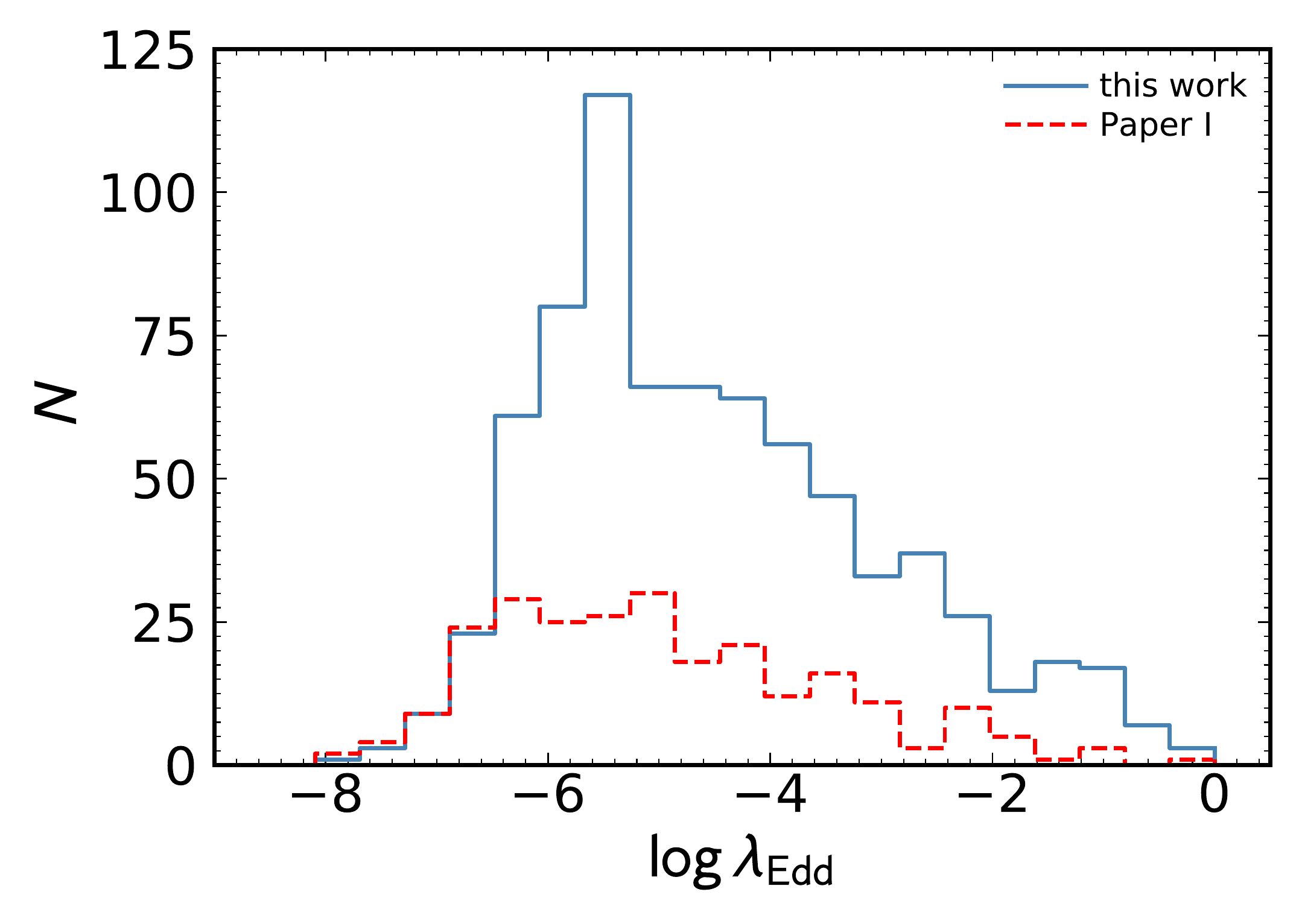}
\caption{Distributions of the X-ray luminosity (\textbf{top}) and Eddington ratio (\textbf{bottom}) for AGN candidates in our sample and in \citetalias{She2017_PaperI}. }
\label{fig:lx_dist}
\end{figure}

Here the new sample is compared with the one in \citetalias{She2017_PaperI}.  The distributions of the distance to the host galaxy is shown in Figure~\ref{fig:dist} for the two samples.  With the update of the Chandra archive and NED, also with the inclusion of the group catalog, 266 new objects in our sample fall into 50~Mpc, while the majority are beyond this distance. The comparison of Hubble types of the host galaxies is shown in Figure~\ref{fig:type}.  As one can see, with a larger distance, the new sample tends to include more early-type galaxies.  The distributions of the X-ray luminosity and the Eddington ratio for AGN candidates are shown in Figure~\ref{fig:lx_dist}. AGN candidates in addition to those in \citetalias{She2017_PaperI} have a higher luminosity on average, which can be understood as a selection effect.  However, due to inclusion of many early-type galaxies, where more massive black holes may lurk, the Eddington ratios of the additional AGN candidates do not occupy the high end but have a fairly similar distribution as previous.  78\% of the new AGN candidates have an Eddington ratio less than $10^{-4}$, constituting a valuable sub-sample for the study of the accretion physics for LLAGNs.

\section{Results}
\label{sec:result}

\begin{deluxetable}{llll}
\tablewidth{0pt}
\tablecaption{Fraction of AGN as a function of Hubble type
\label{tab:type}}
\tablehead{
\colhead{Class} & \colhead{\citetalias{She2017_PaperI}} & \colhead{This work} & \colhead{This work} \\
\colhead{} & \colhead{} & \colhead{(all)} & \colhead{($10^{39}$~\ergs)}}
\startdata
E & $80/142=0.56^{+ 0.06}_{- 0.06}$ & $187/331=0.56^{+0.04}_{-0.04}$ & $46/158 = 0.29^{+0.06}_{-0.05}$ \\
S0 & $70/126=0.55^{+ 0.07}_{- 0.07}$ &  $170/363=0.46^{+0.04}_{-0.04}$ & $56/168 = 0.33^{+0.06}_{-0.05}$ \\
S0/a-Sab & $47/71=0.65^{+ 0.08}_{- 0.09}$ & $111/195=0.56^{+0.05}_{-0.05}$ & $37/82 = 0.45^{+0.08}_{-0.08}$ \\
Sb-Sbc & $53/95=0.55^{+ 0.08}_{- 0.08}$ &  $92/189=0.48^{+0.05}_{-0.05}$ & $43/107 = 0.40^{+0.07}_{-0.07}$ \\
Sc-Scd & $39/100=0.39^{+ 0.08}_{- 0.07}$ &  $61/178= 0.34^{+0.05}_{-0.05}$ & $19/119 = 0.16^{+0.05}_{-0.05}$ \\
Sd-Sdm &  $10/52=0.20^{+ 0.09}_{- 0.08}$ &  $11/64=0.18^{+0.08}_{-0.07}$ & $2/54 = 0.05^{+0.05}_{-0.03}$ \\
Sm-Im & $5/60=0.10^{+ 0.06}_{- 0.05}$ & $10/78=0.13^{+0.06}_{-0.05}$ & $3/68 = 0.05^{+0.05}_{-0.03}$ \\
I0 & $4/6=0.62^{+0.24}_{-0.28}$ & $5/10=0.50^{+0.22}_{-0.22}$ & $2/7 = 0.33^{+0.26}_{-0.22}$ \\
pec & $1/6=0.25^{+0.27}_{-0.19}$ & $8/14=0.56^{+0.19}_{-0.20}$ & $1/8 = 0.20^{+0.22}_{-0.15}$ \\
Unknown & $5/61=0.09^{+0.06}_{-0.05}$ & $127/542=0.23^{+0.03}_{-0.02}$ & $6/108 = 0.06^{+0.04}_{-0.03}$ \\
All &  $314/719=0.43^{+0.03}_{-0.03}$ &  $782/1964=0.39^{+0.01}_{-0.01}$ & $215/879 = 0.24^{+0.02}_{-0.02}$ \\
\enddata
\tablecomments{Errors are quoted at 90\% confidence level. The third column gives the fractions using all of the data, and the fourth column includes data with a sensitivity better than $10^{39}$~\ergs.}
\end{deluxetable}

\begin{deluxetable}{llll}
\tablewidth{0pt}
\tablecaption{Fraction of AGN as a function of optical spectral classification
\label{tab:class}}
\tablehead{
\colhead{Spectral class} & \colhead{\citetalias{She2017_PaperI}}  &  \colhead{This work} & \colhead{This work} \\
\colhead{} & \colhead{} & \colhead{(all)} & \colhead{($10^{39}$~\ergs)}}
\startdata
Seyfert 1 & $18/19 = 0.90^{+0.07}_{-0.12}$ & $27/30=0.87^{+0.07}_{-0.10}$ & $13/13 = 0.93^{+ 0.06}_{- 0.12}$ \\
Seyfert 2 & $34/40 = 0.83^{+ 0.08}_{- 0.10}$ & $55/67=0.81^{+ 0.07}_{- 0.08}$ & $23/39 = 0.58^{+ 0.12}_{- 0.12}$ \\
Seyfert 1 and 2 & $52/59 = 0.86^{+ 0.06}_{- 0.07}$ & $82/97=0.83^{+ 0.05}_{- 0.06}$ & $36/52 = 0.68^{+ 0.09}_{- 0.10}$ \\
LINER 1 & $16/16 = 0.94^{+ 0.05}_{- 0.10}$ & $20/20=0.95^{+0.04}_{- 0.08}$ & $14/17 = 0.78^{+ 0.13}_{- 0.16}$ \\
LINER 2 & $41/50 = 0.80^{+ 0.08}_{- 0.09}$ &  $53/65=0.80^{+0.07}_{-0.08}$ & $24/46 = 0.52^{+ 0.11}_{- 0.11}$ \\
LINER 1 and 2 & $57/66 = 0.85^{+ 0.06}_{- 0.07}$ &  $73/85=0.85^{+0.05}_{-0.06}$ & $38/63 = 0.60^{+ 0.09}_{- 0.10}$ \\
transition & $28/41 = 0.67^{+ 0.11}_{- 0.12}$ &  $30/46=0.64^{+0.10}_{-0.11}$ & $17/40 = 0.42^{+ 0.12}_{- 0.12}$ \\
\HII\ & $51/163 = 0.31^{+ 0.06}_{- 0.05}$ & $51/180=0.28^{+0.05}_{-0.05}$ & $19/163 = 0.12^{+ 0.04}_{- 0.03}$ \\
absorption-line & $55/89 = 0.61^{+ 0.08}_{- 0.08}$ & $56/96=0.58^{+0.08}_{-0.08}$ & $24/86 = 0.28^{+ 0.08}_{- 0.07}$ \\
All & $243/418 = 0.58^{+ 0.03}_{- 0.03}$ & $292/504=0.57^{+0.03}_{-0.03}$ & $134/404 = 0.33^{+ 0.03}_{- 0.03}$ \\
\enddata
\tablecomments{Errors are quoted at 90\% confidence level. The third column gives the fractions using all of the data, and the fourth column includes data with a sensitivity better than $10^{39}$~\ergs.}
\end{deluxetable}

Compared with \citetalias{She2017_PaperI}, the galaxy sample size expands from 719 to 1,964, and the number of AGN candidates increases from 314 to 782 in the new sample. This allows us to repeat the experiments based on the sample in \citetalias{She2017_PaperI} to a better precision. AGN fractions are listed in Table~\ref{tab:type} as a function of the Hubble type and in Table~\ref{tab:class} as a function of optical classification.  In the paper, the AGN fraction is calculated using the Bayesian inference of binomial proportion assuming a uniform prior between 0 and 1. The estimate of the fraction is thus $\left(n_{\rm AGN} + 1\right) / \left(n_{\rm gal} + 2\right)$.  The errors are quoted as the equal-tailed interval of the posterior. \citetalias{She2017_PaperI} did not take into account the effect of uneven sensitivities in luminosity. We add columns in Table~\ref{tab:type} \& \ref{tab:class} for observations with $L_{\rm limit} < 10^{39}$~\ergs\ and accordingly AGNs in them with $L_{\rm X} > 10^{39}$~\ergs.

\begin{figure}
\centering
\includegraphics[width=\columnwidth]{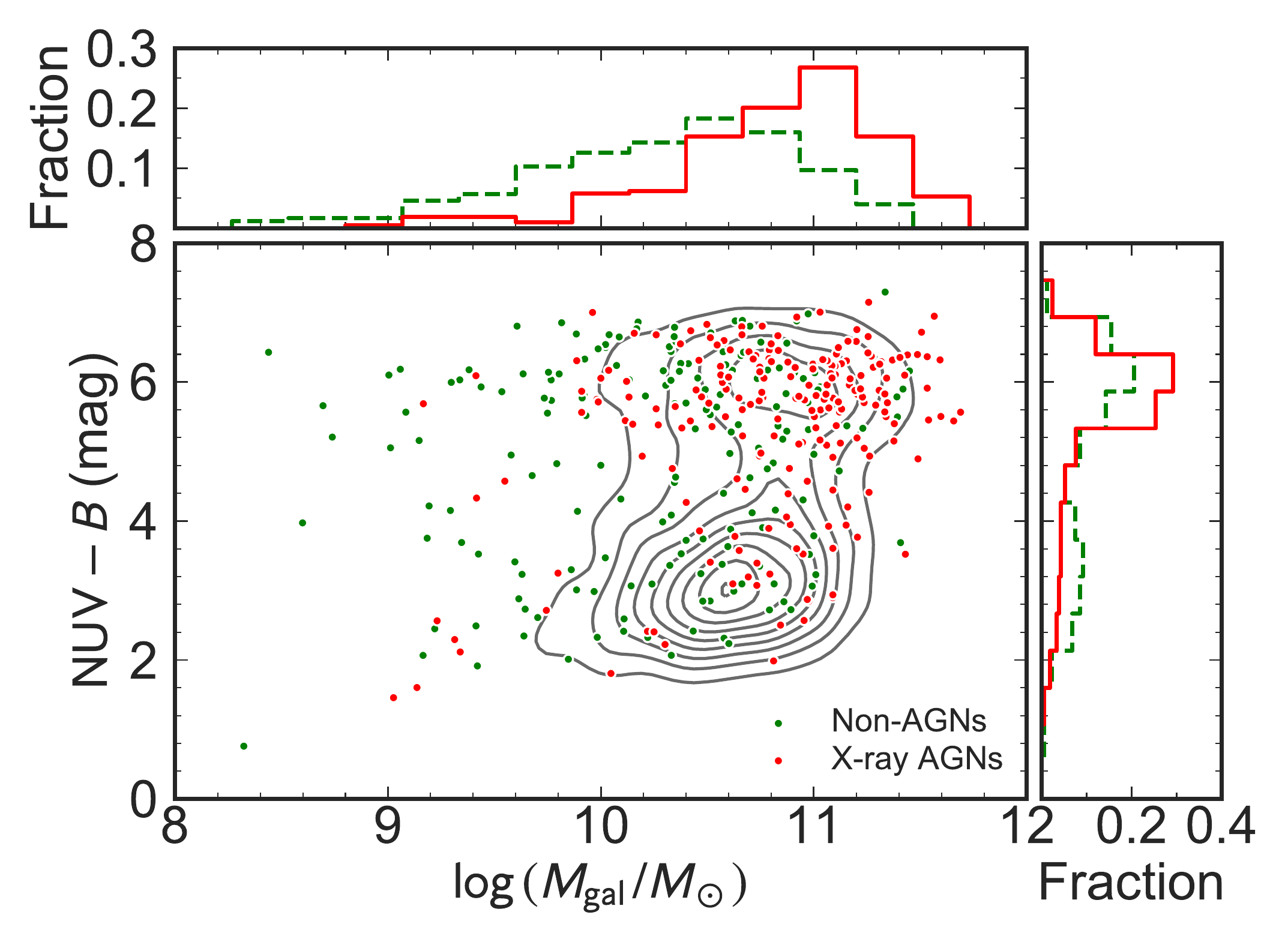}\\
\caption{Color-mass diagram for galaxies in our sample. The contours are created using all of the galaxies in the \citet{Tully2015} catalog.}
\label{fig:color_mass}
\end{figure}

\begin{figure}[t]
\centering
\includegraphics[width=\columnwidth]{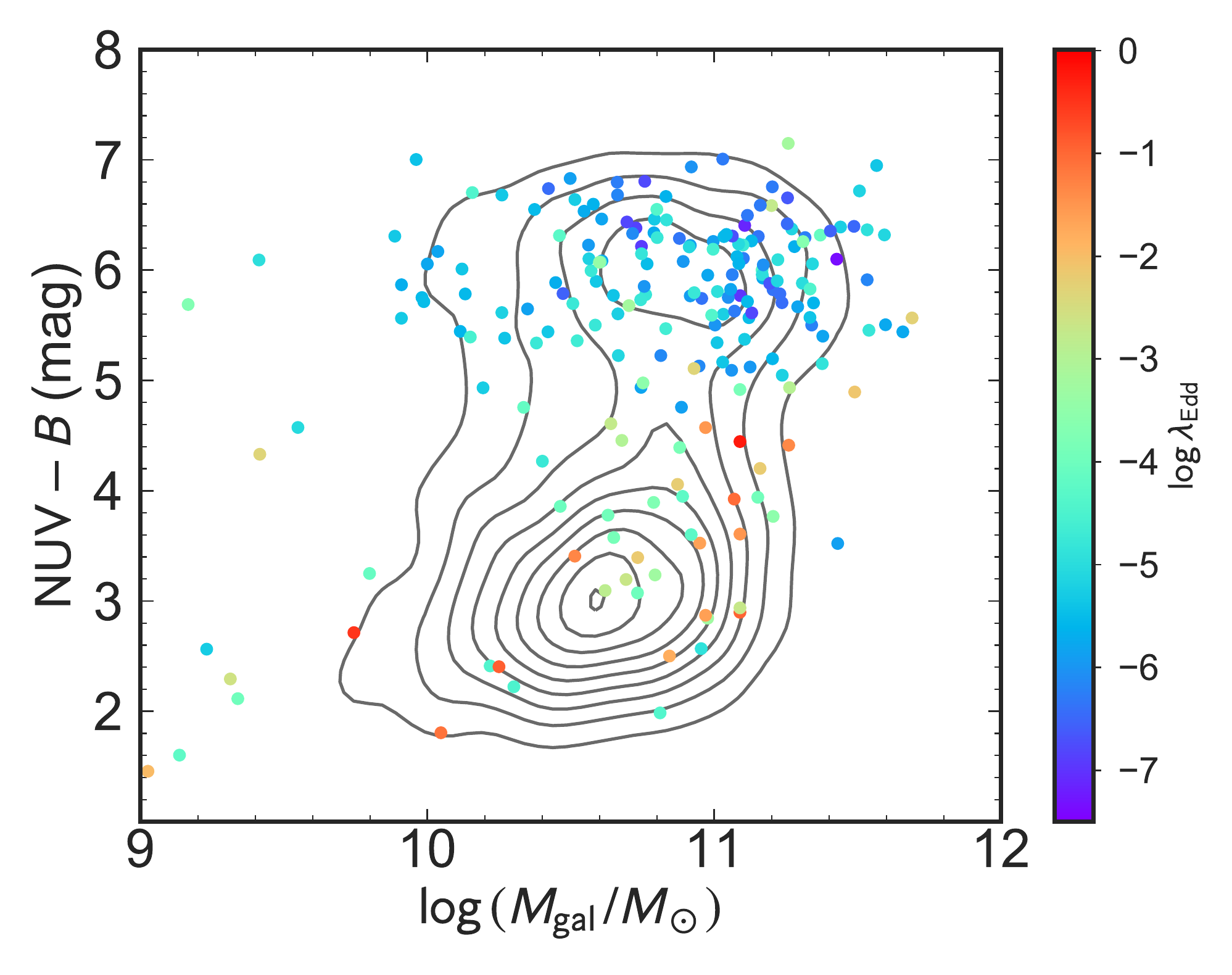}\\
\caption{X-ray AGNs on the color-mass diagram with Eddington ratios indicated by the colors on the color bar.  The contours are created using all of the galaxies in the \citet{Tully2015} catalog. }
\label{fig:color_mass_eddr}
\end{figure}

\begin{figure}[t]
\centering
\includegraphics[width=\columnwidth]{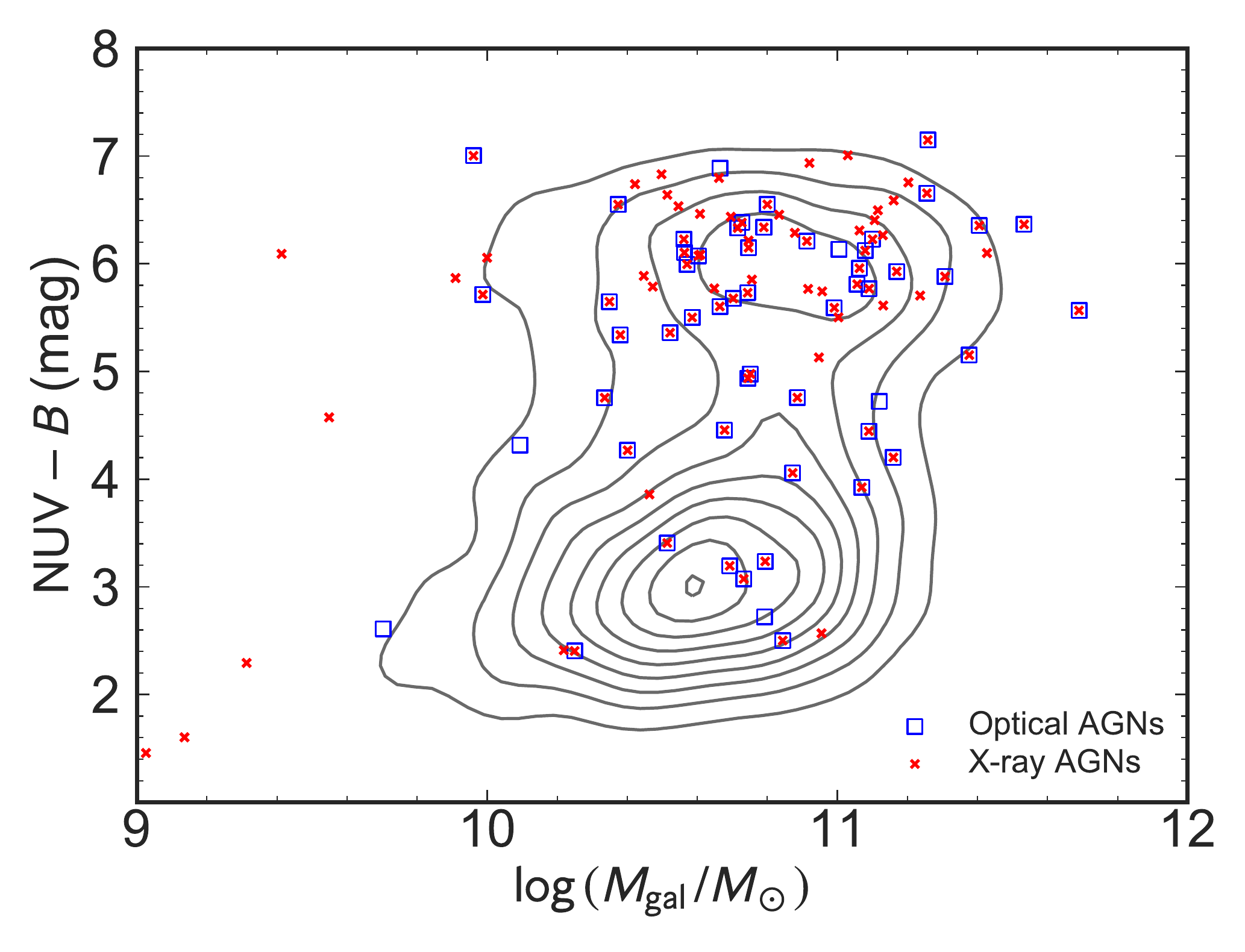}\\
\caption{AGNs selected in optical and X-ray on top of the color-mass diagram.  The contours are created using all of the galaxies in the \citet{Tully2015} catalog.}
\label{fig:color_mass_xo}
\end{figure}

\begin{figure}[t]
\centering
\includegraphics[width=\columnwidth]{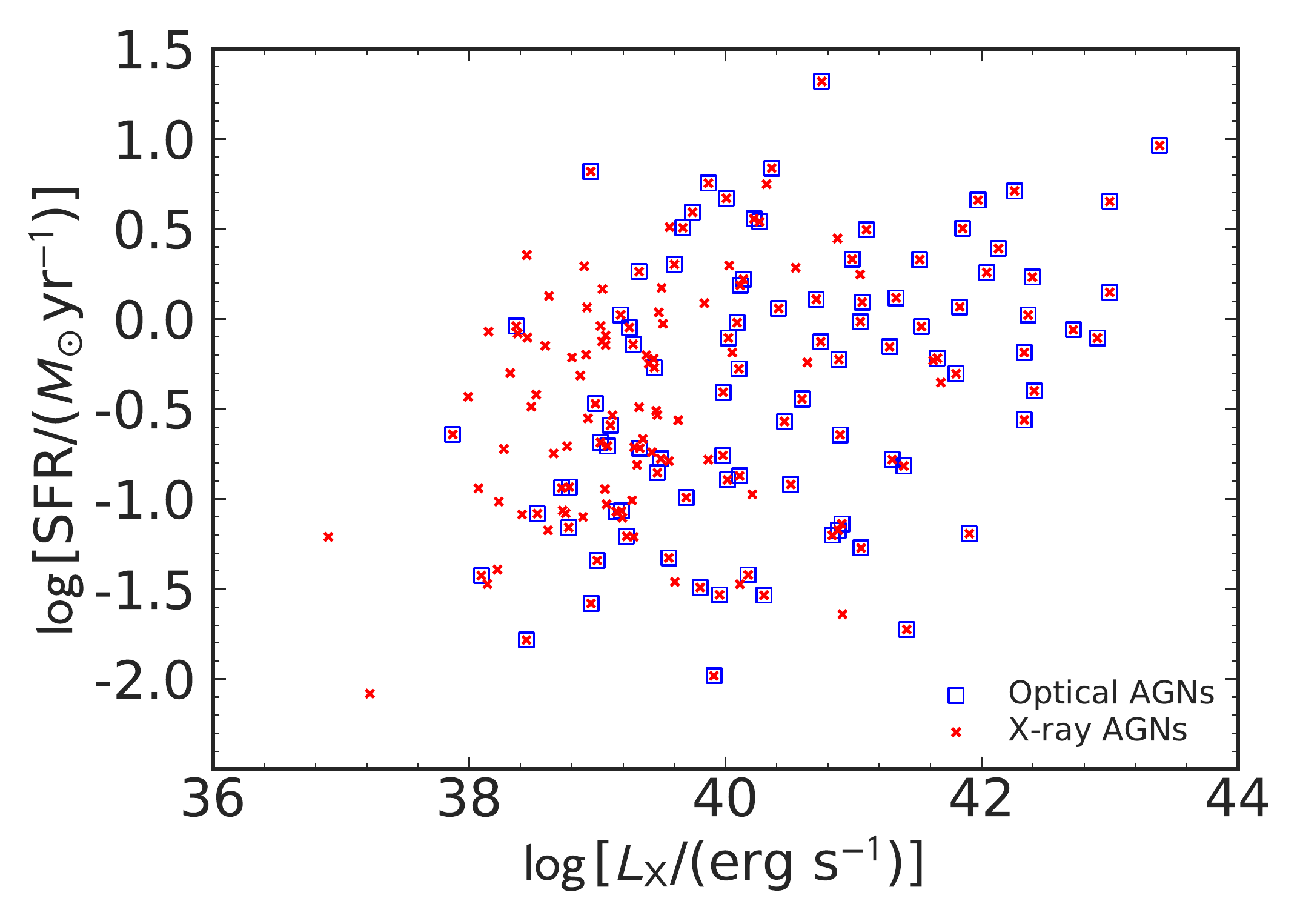}\\
\caption{SFR vs.\ $L_{\rm X}$ for AGNs selected in optical and X-ray. }
\label{fig:sfr_lx_xo}
\end{figure}

We plot the color-mass diagram in Figure~\ref{fig:color_mass}, for both AGN and non-AGN host galaxies in our sample. The background contours are created using all of the galaxies in the \citet{Tully2015} catalog, to outline the distributions of the red sequence, green valley, and blue cloud \citep{Bell2004}. It is obvious that our sample is more toward the red sequence, containing a large fraction of less active, elliptical galaxies.  The detection rate of X-ray AGNs is also more pronounced in the red sequence, where the AGN Eddington ratio is much lower (Figure~\ref{fig:color_mass_eddr}).  AGNs in our sample that can be detected in X-ray but not in optical also cluster in the red sequence (Figure~\ref{fig:color_mass_xo}), suggesting that their absence in the optical search is not due to stellar contamination. On the other hand, the majority (at least 80\%, see Table~\ref{tab:class}) of the optically selected AGNs, except those with exceptionally low-sensitivity Chandra observations, can be picked up by the X-rays.  The SFR-$L_{\rm X}$ diagram (Figure~\ref{fig:sfr_lx_xo}) suggests that, compared with optical, X-rays are more sensitive to select AGNs whose absolute activity ($L_{\rm X}$) is low.  To conclude, compared with the optical approach, X-rays indeed pick up intrinsically low-activity AGNs, while the optical AGN classifications are largely robust. 

AGNs in our sample that can be detected in X-ray but not in optical also cluster in the red sequence with low Eddington ratios (Figure~\ref{fig:color_mass_xo}), which indicates that the majority of the optically selected AGNs can be picked up by X-ray observations (at least $\sim$80\%, see Table~\ref{tab:class}), with the exceptions mainly due to sensitivity issues.  
With the new sample, to summarize, all of the conclusions based on the previous sample remain valid. Here we mention the two most interesting results, one about the AGN or black hole occupation fraction in late-type galaxies, in particular, those with an \HII\ nucleus, and the other about the accretion physics for LLAGNs.

The AGN fraction sets a lower limit on the black hole occupation fraction.  The main science drive for \citetalias{She2017_PaperII} is to constrain the occupation fraction of central black holes via AGN activities in late-type, bulgeless galaxies, which can be used to test the formation mechanism of supermassive black holes in the early universe.  Here, with the updated sample, we find an AGN fraction of 28\% $\left[= \left(51+1\right) / \left(180+2\right)\right]$ in \HII\ nuclei, consistent with the value found in \citetalias{She2017_PaperII}. Most likely, these galaxies contain no or very small bulges, and consequently low-mass black holes that have not been through much evolution since their formation. Following the argument in \citetalias{She2017_PaperII} that 26\% of the AGN candidates with a luminosity above $10^{38}$~\ergs\ may be contaminated by X-ray binaries in the nuclear star cluster, we obtain a lower limit of 19\% for low-mass black holes in late-type galaxies, considering that there are 46 of the AGN candidates in \HII\ nuclei above $10^{38}$~\ergs. If we select observations with $L_{\rm limit} < 10^{39}$~\ergs, and accordingly, AGNs in them with $L_{\rm X} > 10^{39}$~\ergs, assuming a contamination of 8\% from X-ray binaries \citepalias{She2017_PaperII}, the fraction is 11\%~$\left[= 0.92 \times \left(19+1\right) / \left(163+2\right)\right]$. 

\begin{figure}[t]
\centering
\includegraphics[width=\columnwidth]{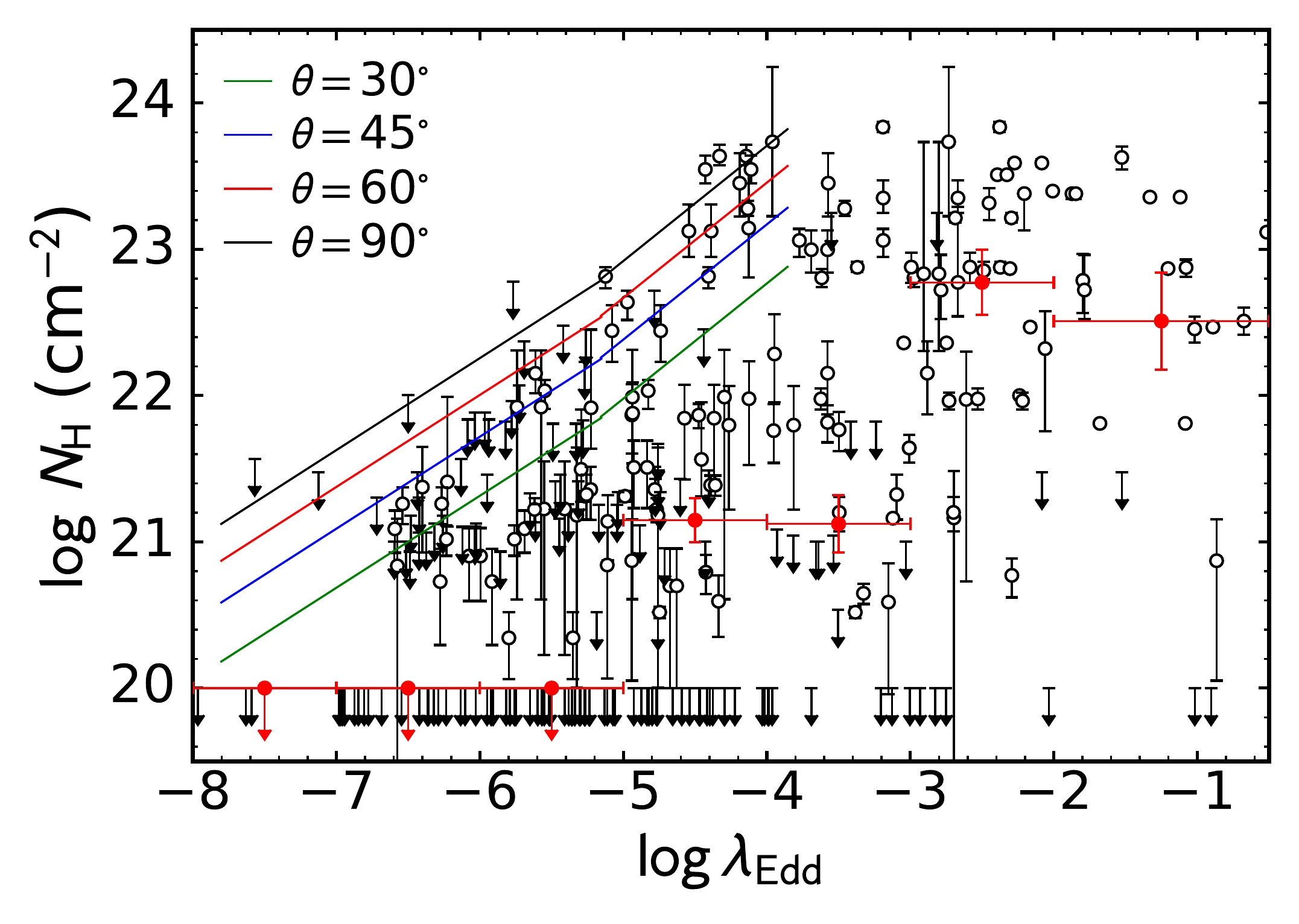}
\includegraphics[width=\columnwidth]{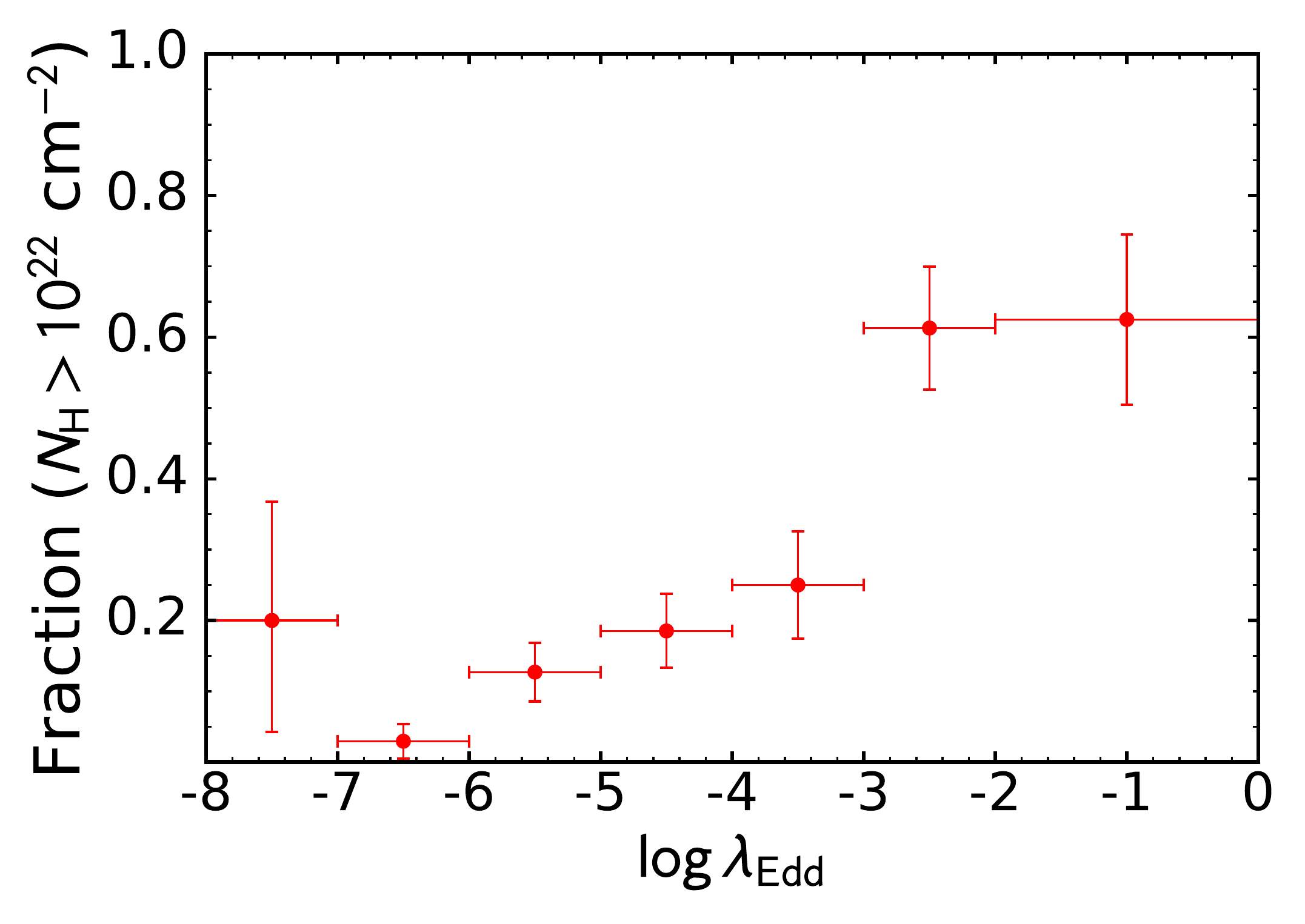}
\caption{Intrinsic absorption column density $N_{\rm H}$ (\textbf{top}) and the fraction of highly absorbed AGNs (\textbf{bottom}) as a function of the Eddington ratio. The red points are the median values in each bin. In the top panel, the solid lines indicate the hydrogen column density of the outflow at different viewing angles, adopted from numerical simulations of the hot accretion flow \citep{Yuan2015}.}
\label{fig:nh_vs_edd}
\end{figure}

LLAGNs are of particular interest as they are ideal sites for the study of the hot accretion flow. In \citetalias{She2018_PaperIII}, \citet{She2018_PaperIII} reported the discovery that either the absorption column density or the fraction of highly absorbed AGNs is scaled with the Eddington ratio. Thanks to the increase of sample size at Eddington ratios below $10^{-4}$, as well as the choice of a new mass estimator, the findings in \citetalias{She2018_PaperIII} can be tested to a better precision.  In Figure~\ref{fig:nh_vs_edd}, we plot the intrinsic (beyond the Milky Way) X-ray absorption column density $N_{\rm H}$ or the fraction of objects with $N_{\rm H} > 10^{22}$~cm$^{-1}$ vs.\ the Eddington ratio. The positive correlation is still seen, with smaller error bars.

In addition to an expansion in distance and consequently in the sample size, 1,557 objects in this sample are also contained in the galaxy group catalog of \citet{Tully2015}, offering us environmental information needed for the study of AGN triggering and feeding.  Here we display the AGN fraction as a function of galaxy morphology, group size, and host position in the group (central or satellite). As the distance distributions for early and late-type galaxies are not the same, we, again, select observations with $L_{\rm limit} < 10^{39}$~\ergs\ and AGN candidates with $L_{\rm X} > 10^{39}$~\ergs\ for this study,  to avoid selection effect due to uneven sensitivity. 

\begin{figure}
\centering
\includegraphics[width=\columnwidth]{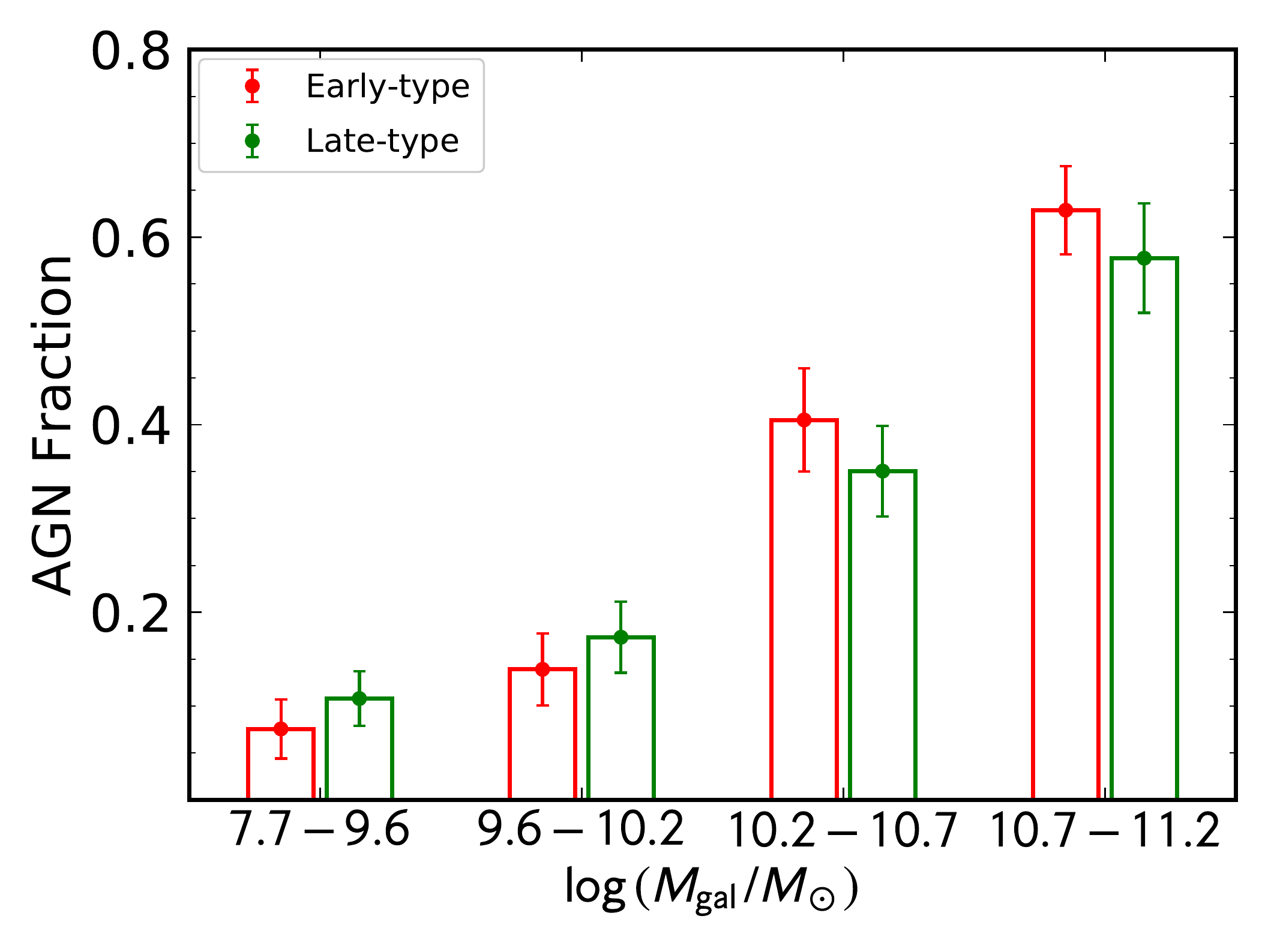}
\caption{AGN fraction as a function of galaxy stellar mass for early and late-type galaxies.}
\label{fig:el_mass}
\end{figure}

The AGN fraction in early-type (E and S0) galaxies vs.\ that in late-type (S and later) galaxies are shown in Figure~\ref{fig:el_mass}.  The comparison is controlled by the mass of the host galaxy.  Four mass bins are used, with approximately the same number of objects in each.  The errors are quoted at the 68\% credible interval.

\begin{figure}
\centering
\includegraphics[width=\columnwidth]{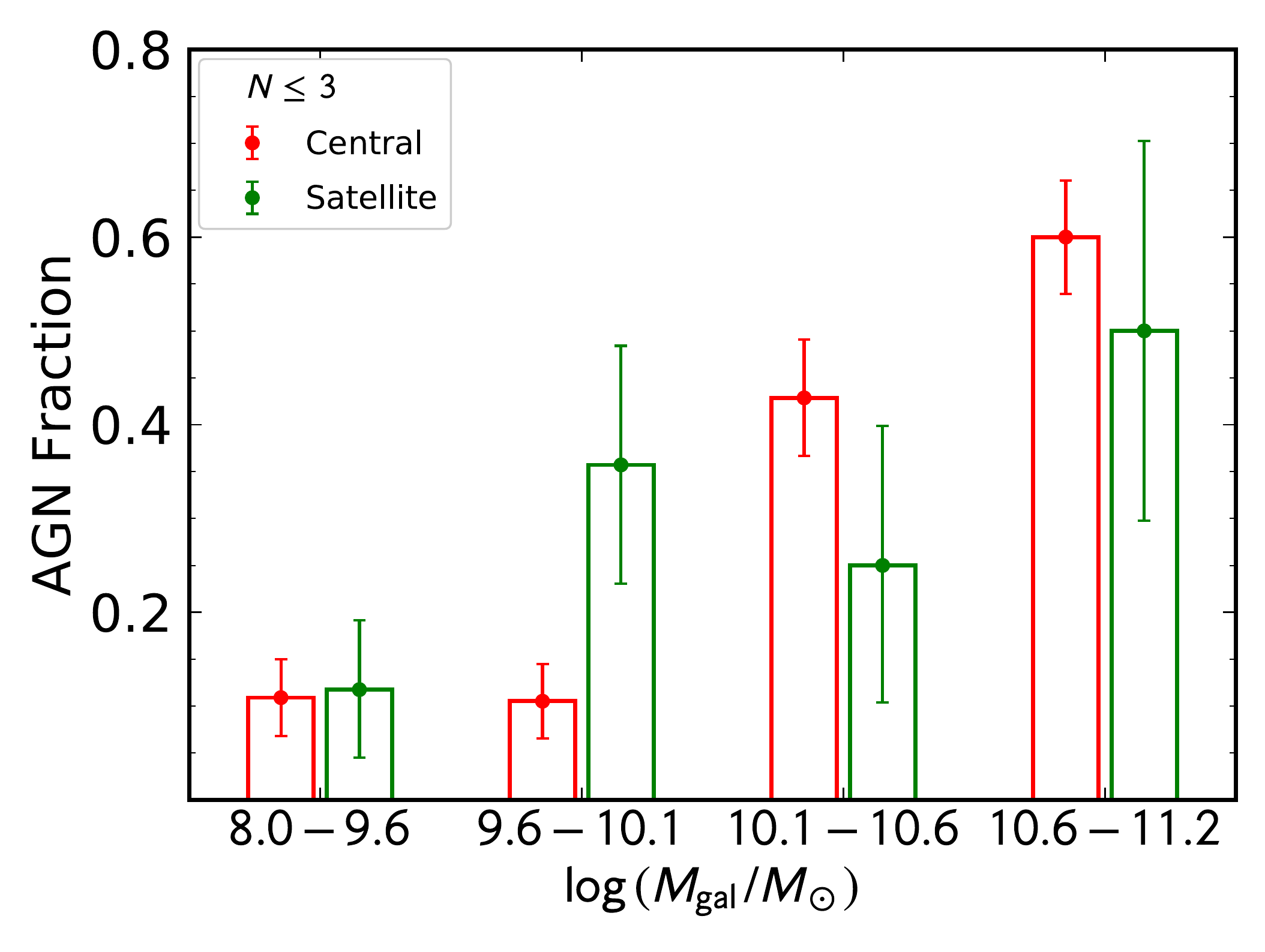}
\includegraphics[width=\columnwidth]{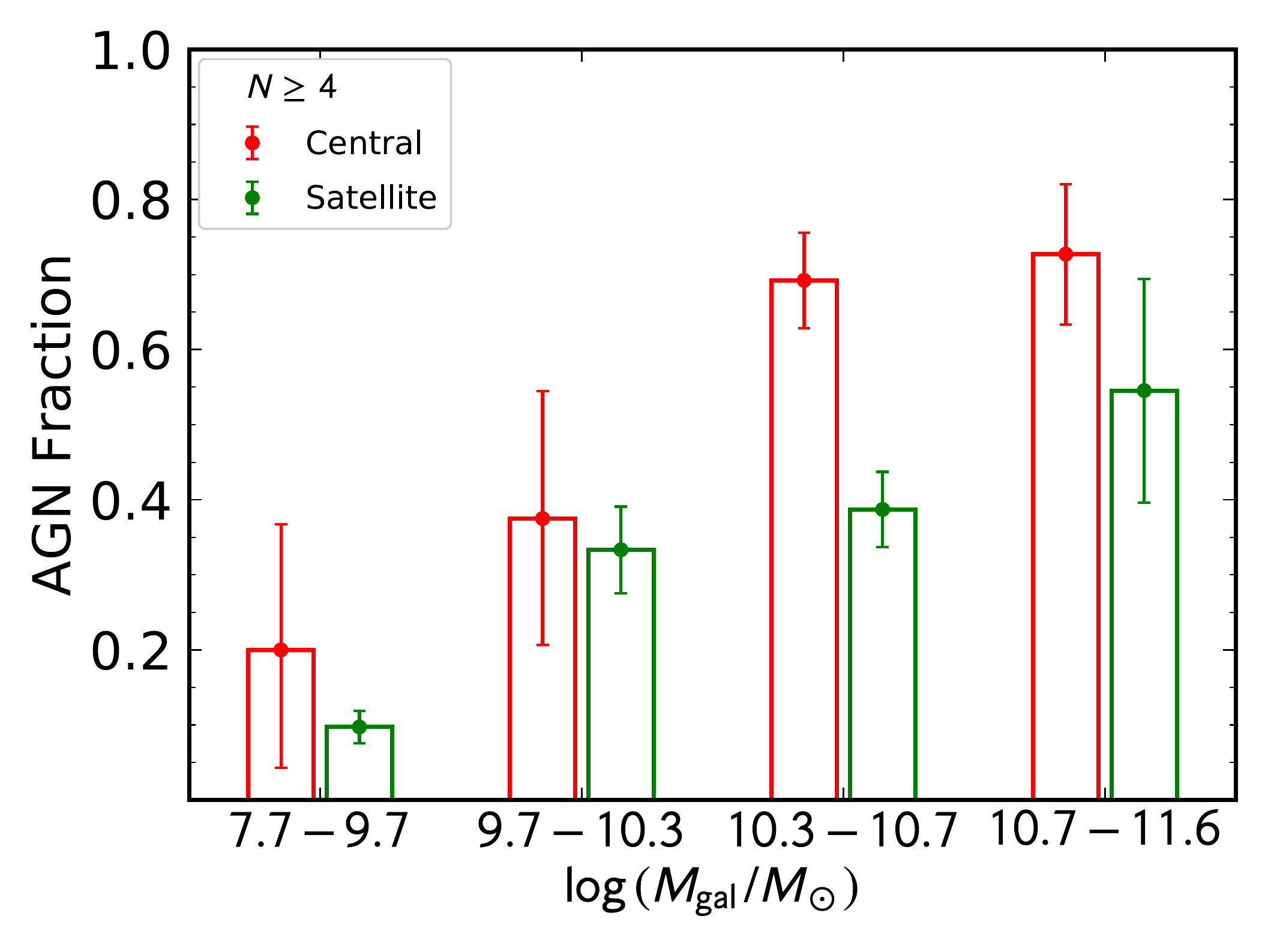}
\caption{AGN fraction as a function of galaxy stellar mass for central and satellite galaxies in small ($N \le 3$; \textbf{top}) and large ($N \ge 4$; \textbf{bottom}) galaxy groups. }
\label{fig:cs_mass}
\end{figure}

\begin{figure}[htbp]
\centering
\includegraphics[width=\columnwidth]{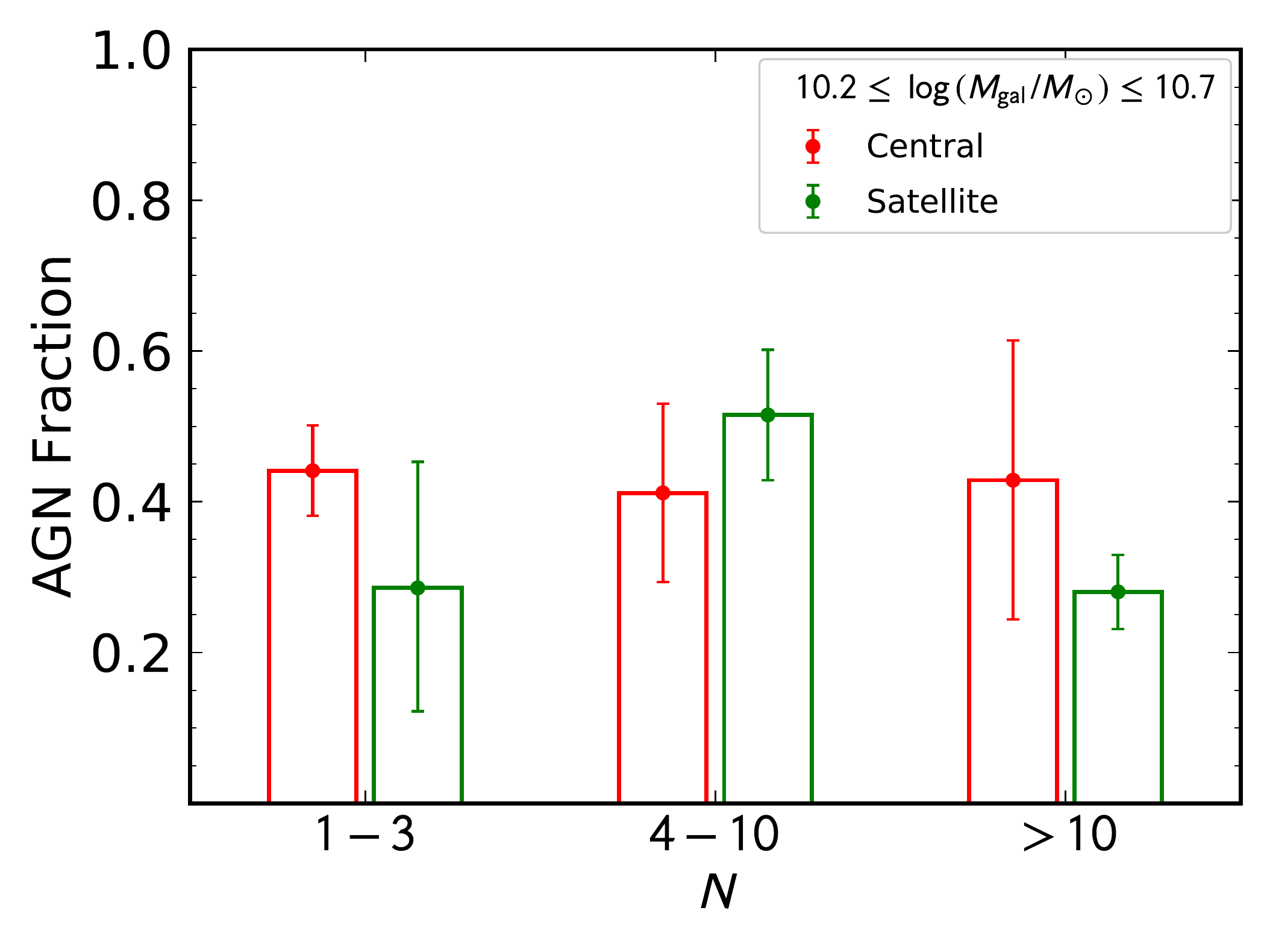}
\includegraphics[width=\columnwidth]{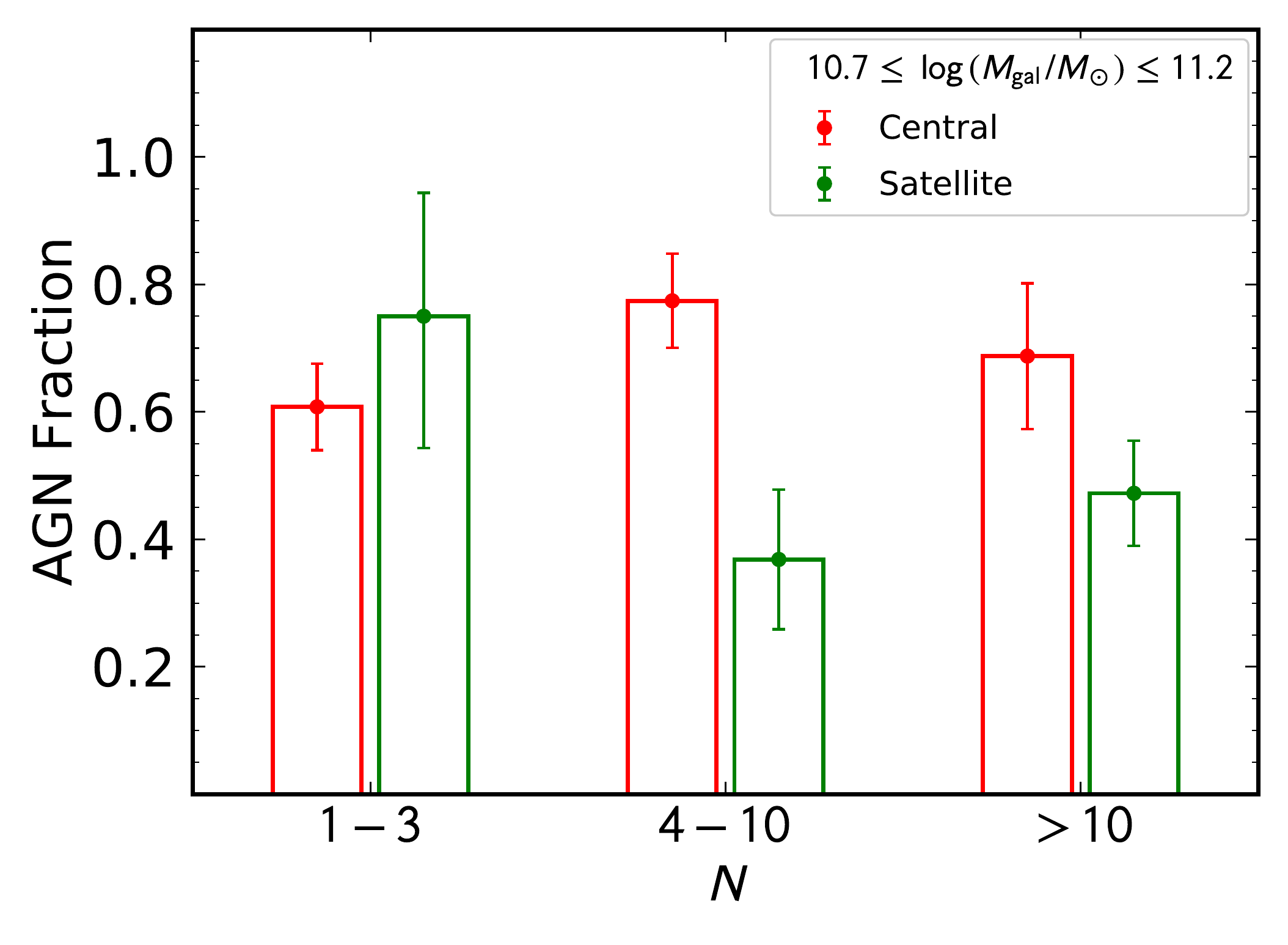}
\caption{AGN fraction as a function of group size for central and satellite galaxies in two mass bins.}
\label{fig:cs_size}
\end{figure}

Similarly, the AGN fraction for central and satellite galaxies is compared in Figure~\ref{fig:cs_mass} as a function of the host mass in two group size bins ($N \le 3$ and $N \ge 4$), and in Figure~\ref{fig:cs_size} as a function of group size in two mass bins ($10^{10.2 - 10.7}$~$M_\sun$ and $10^{10.7 - 11.2}$~$M_\sun$). 

\section{Discussions}
\label{sec:discussion}

In this paper, we expand the search of X-ray AGNs in nearby galaxies in the Chandra archive to a distance of 150~Mpc, and obtain a larger sample. The conclusions about the AGN occupation fraction in \HII\ nuclei \citepalias{She2017_PaperII}, and about the correlation between the absorption and Eddington ratio \citepalias{She2018_PaperIII}, are further confirmed.  As these two topics have been extensively discussed in  \citetalias{She2017_PaperII} and \citetalias{She2018_PaperIII}, respectively, we will not repeat the discussions here, but focus on the physical constraints on AGN triggering and feeding. 

The secular process may be the dominant mechanism to feed low-mass SMBHs in late-type, gas-rich galaxies \citep{Hopkins2014}, occurring on evolutionary timescales.  A remarkable example could be the narrow-line Seyfert 1 galaxies \citep{OrbandeXivry2011}.  On the other hand, black holes in group center galaxies could be fed by cold gas streams in the intergalactic medium, the so-called chaotic cold accretion, which is suggested by theoretical models and simulations \citep{Pizzolato2005,Li2014} and supported by detections of molecular gases \citep{David2014,Temi2018} and H$\alpha$ filaments \citep{Fabian2012}.  The chaotic cold accretion could be significant in groups of large sizes \citep{Gaspari2015,Gaspari2017,Gaspari2018}.  Similar conceptions include the precipitation \citep{Voit2015} and stimulated feedback \citep{Mcnamara2016}, with the same idea that efficient accretion can take place only if the hot ambient medium condenses into cooler clouds due to thermally unstable cooling \citep{Werner2019}.

To test the secular process, one may expect that the AGN fraction in late-type galaxies is higher than in early-types. In Figure~\ref{fig:el_mass}, there is no significant difference between the fractions in the two types, which may suggest that the secular process is not the dominant mechanism that triggers and feeds LLAGNs in nearby galaxies. The reason that it does not take a leading role in these cases is probably due to the fact that most of the AGNs in our sample are LLANGs, while the secular process may trigger high-luminosity AGNs \citep{OrbandeXivry2011}.  It is consistent with the fact that the bars have no effect on the nuclear activity \citep{Ho1997}. 

To test the thermally unstable cooling scenario, one may expect that central galaxies, especially in large groups, are more likely to be triggered.  Again, we do not see any significant difference in the AGN fraction between central and satellite galaxies (Figure~\ref{fig:cs_mass}), or an increasing AGN fraction with increasing group size (Figure~\ref{fig:cs_size}). Although the null results tend to rule out the thermally unstable cooling being the dominant process in triggering LLAGNs, such a conclusion may be considered with cautions, as the central and satellite AGNs do show differences. For examples, the radio mode is more prevalent in the former \citep{Fabian2012}, and some massive early-type satellites (e.g., those in the Virgo cluster) may harbor their own atmospheres to feed the AGNs.

\acknowledgements We thank the anonymous referee for useful comments that have helped improve the interpretation of the results. HF acknowledges funding support from the National Key R\&D Project (grants Nos.\ 2018YFA0404502 \& 2016YFA040080X), and the National Natural Science Foundation of China under the grant Nos.\ 11633003 \& 11821303. LCH was supported by the National Science Foundation of China (11721303, 11991052) and the National Key R\&D Program of China (2016YFA0400702).


\end{document}